\documentclass{desyproc}
\usepackage{paralist}
\usepackage{subfig}
\usepackage{array}
\usepackage[colorlinks=true, citecolor=black, urlcolor=black, linkcolor=black]{hyperref}

\begin{document}
\title{Status report of the CERN microwave axion experiment}

\author{{\slshape M. Betz, F. Caspers, M. Gasior}\\[1ex]
European Organization for Nuclear Research (CERN), Geneve, Switzerland\\
}

\contribID{Betz\_Michael}

\desyproc{DESY-PROC-2013-XX}
\acronym{Patras 2013} 
\doi  
\maketitle

\begin{abstract}
``Light Shining Through the Wall'' experiments can probe the existence of ``axion like particles'' through their weak coupling to photons. We have adapted such an experiment to the microwave regime and constructed the table top apparatus. This work presents an overview of the experimental setup and then focuses on our latest measurement run and its results.
By operating the apparatus within a superconducting MRI magnet, competitive exclusion limits for axion like particles to the first generation optical light shining through the wall experiments have been achieved.  
\end{abstract}

\section{Introduction}
\begin{figure}[tbh]
\begin{center}
\subfloat[Simplified schematic of the experimental setup]{
\includegraphics[width=0.55\textwidth]{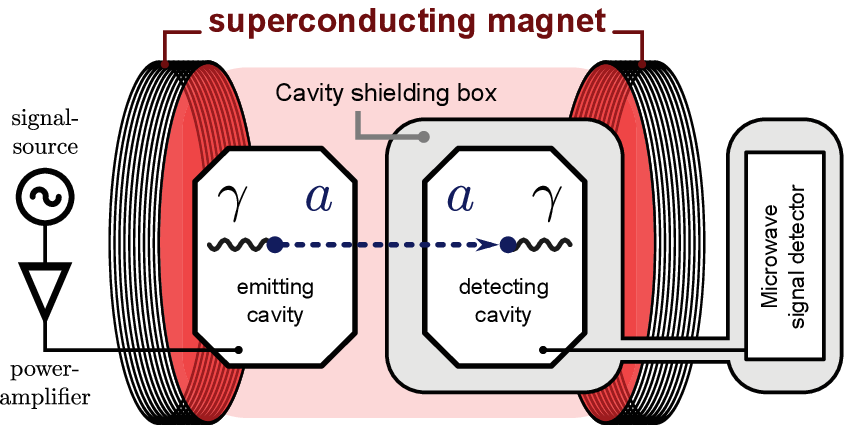}
\label{fig:ovrBlock}
}
\subfloat[Emitting cavity \& detecting cavity in shielding box]{
\includegraphics[width=0.4\textwidth]{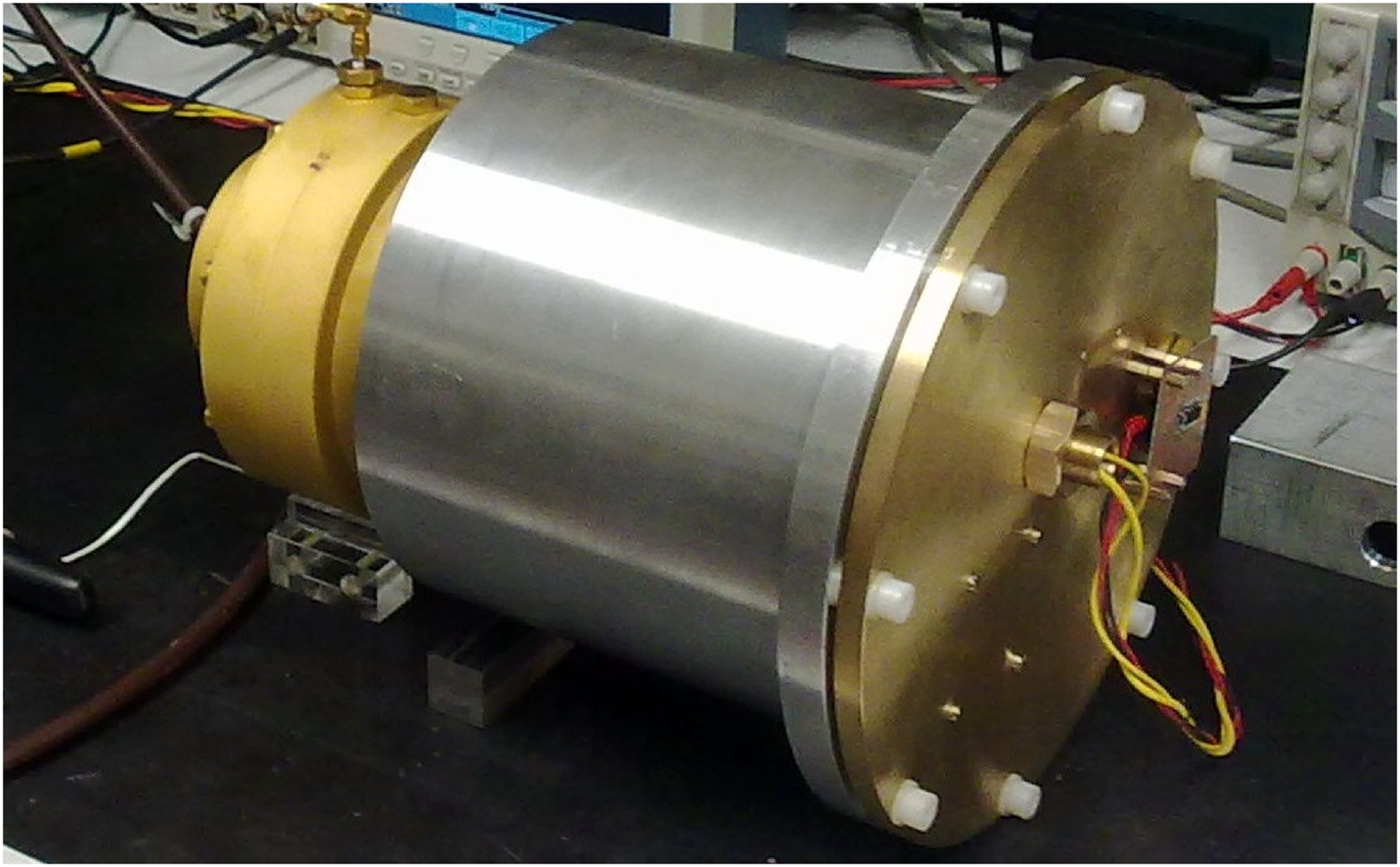}
\label{fig:ovrPhoto}
}
\end{center}
\caption{Overview of the light shining through the wall experiment in the microwave range configured to search for ALPs}
\label{fig:theExp}
\end{figure}

The concept of an optical Light Shining Through the Wall (LSW) experiment has been adapted to microwaves as described in \cite{src:JaCaRi}.
A block diagram of the setup at CERN as it has been used to search for Axion Like Particles (ALPs) with microwaves
 is shown in Fig.~\ref{fig:ovrBlock}, it consists of two identical microwave cavities with a loaded quality factor of $Q \approx 12000$ and a spacing between them of less than 20~mm.
One serves as ALP emitter and is excited by 50~W of RF power on its resonant frequency $f_{\mathrm{sys}} = 1.739990$~GHz. It develops a strong electromagnetic (EM) field, corresponding to a large number of microwave photons $\gamma$.
These can convert to ALPs by the Primakoff effect \cite{PhysRevLett.51.1415}. ALPs do not interact with matter and propagate towards the detection cavity, which is connected to a very sensitive microwave receiver. The reciprocal
conversion process transforms ALPs back to microwave photons, which can be observed as an excitation of
the seemingly empty and well shielded detection cavity.
Since there is no energy loss associated with the ALP conversion process, the signal from the detection cavity would be observable at exactly the frequency $f_{sys}$, making a narrowband receiving concept feasible.

\section{Engineering challenges}
In this section, we give an overview of the critical engineering challenges, which were encountered during the realization of the microwave based LSW experiment:
\begin{description}
\item[Electromagnetic shielding]
Shielding is required around the detecting cavity and the microwave receiver to eliminate ambient
electromagnetic interference (EMI) and to mitigate coupling to the emitting cavity by electromagnetic
leakage. This would generate false positive results, as a signal originating
from leakage can not be discriminated against an ALP signal. Within 2 cm,
the field strength must be reduced by at least a factor of $7.7 \cdot 10^{12} = 258 \mathrm{~dB}$ to
obtain meaningful results and avoid degradation of ALP sensitivity. 
Most microwave components used in the setup like SMA connectors or semi-rigid coaxial cables provide less than 120 dB of shielding, making an external shielding enclosure and strategic use of optical fibres for signal transmission necessary.
With the current setup, over 300 dB shielding effectiveness has been achieved \cite{src:moiIPAC}.

\item[Detection of weak signals]
The smallest detectable signal power during the ALPs run in June 2013 is $P_{det} = -212$~dBm~$= 6.3 \cdot 10^{-25}$~W~$\approx 0.5$~photons/s.
As the experiment is carried out at room temperature, the thermal noise \textbf{density} is $P_{n} \approx -174$~dBm/Hz. A very narrow band filter can be implemented to detect the sinusoidal signal within the background noise.
It is realized by a discrete Fourier transformation over the entire recorded time trace of length $\tau =~10$~h.
Each resulting spectral bin will respond to signals within its resolution bandwidth, given by $\mathrm{BW_{res}} = 1/\tau$.
As noise power scales linearly with resolution bandwidth and signal power stays constant, we achieve a signal to noise ratio improvement proportional to measurement time.
To keep the filter on the right center frequency, a global 10~MHz reference clock is used to phase lock all oscillators involved in the downmixing chain. This detection method has been successfully demonstrated with resolution bandwidths down to 10~$\mu$Hz in \cite{src:narrowband}.      

\item[Keeping the cavities on tune]
For an exclusion result, it is necessary to prove that the detector is working and that a potential ALP signal could not have been concealed.
Detection sensitivity will be limited if the resonant frequency of any of the cavities does not
equal the system frequency $f_{\mathrm{sys}}$ within its 3~dB bandwidth.
The resonant frequency of the cavities can drift due to thermal expansion. For the emitting cavity, the reflected RF power was monitored to ensure it is on resonance during the whole measurement run. For the detecting cavity, its thermal noise density was measured before and after the recording of experimental data, indicating its resonant frequency.

\item[Compatibility with magnetic fields]
As the shielding enclosure and the cavities have to be placed in a strong magnetic field for ALP measurements, they need to be constructed from non-ferromagnetic materials like aluminium or brass. This is to prevent field distortion and to avoid strong attractive forces during insertion of the setup into the magnet.
Some electronic components like the low noise amplifier and the analog optical transmitter need to be placed as close as possible to the detection cavity. Therefore special precautions had to be taken to ensure they operate reliably within the magnetic field. For example, ferrite or iron cored inductors or transformers had to be avoided as the material saturates and changes its magnetic properties in strong magnetic fields.   
\end{description}
\section{Measurement run in June 2013}
\begin{wrapfigure}{r}{0.5\textwidth}
\label{fig:exclSignal}
\vspace{-0.7cm}
\begin{center}
\includegraphics[width=0.5\textwidth]{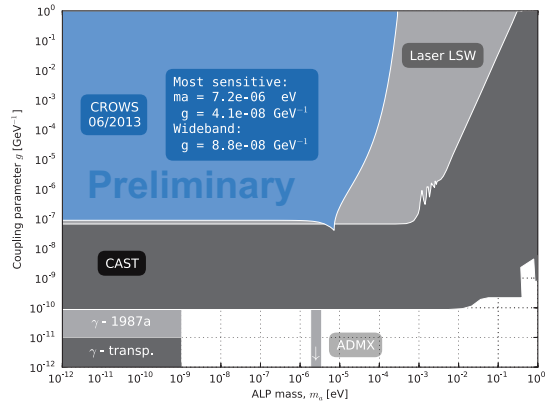}
\caption{Preliminary exclusion limits for ALPs from the CROWS experiment run in June 2013.}
\end{center}
\vspace{-0.5cm}
\end{wrapfigure}
The most sensitive measurement run for ALPs has been carried out in June 2013 in cooperation with the Brain \& Behaviour Laboratory of Geneva University. We were able to operate the setup within the bore of a superconducting magnet, which is part of an MRI scanner. It provides a solenoid-like field of $B = 2.88$~T. Over the course of one weekend, 2~x~10~h of experimental data were recorded.
As no ALPs were detected, the corresponding exclusion limits in comparison to other experiments are shown in Fig.~2.

For diagnostic purposes, a sinusoidal signal of known frequency is emitted within the shielding enclosure. This ``test tone'' of relatively low and constant power ($\approx -100$~dBm) couples from a $\lambda / 4$ antenna to the detection cavity and to the components of the receiver frontend. By identifying the signal in the recorded spectrum, we demonstrate that the entire signal processing chain was operational during the measurement. This also allows to evaluate unwanted frequency offsets, frequency drifts, or phase noise by comparing shape and position of the measured signal peak to the expected one. Furthermore, as long as the test tone signal is observed with constant power during the measurement run, it qualifies that the EM shielding performance has not degraded.

Figure~3 shows the resulting power spectrum from the measurement run. The test tone signal is visible as a single peak, clearly above the noise floor, spanning only one single bin. A frequency window (shaded green) has been defined with a width of $10 \cdot \mathrm{BW_{res}}$, around the frequency where an ALP signal would be expected. 
The peaks within this window do not exceed the detection threshold of $\mathrm{P_{det}} = -212.0~\mathrm{dBm} = 6.3\cdot10^{-25}$~W, allowing us to set an exclusion limit for ALPs.

In order to define the detection threshold, the histogram of $14\cdot10^6$ frequency bins -- containing exclusively spectral background noise -- was evaluated. $\mathrm{P_{det}}$ was set such, that only in 1\% of all measurement runs, a single peak signal above the detection threshold would appear within the WISP window, resulting in a false positive outcome of the experiment.
\begin{figure}[tbh]
\label{fig:resultSpect}
\begin{center}
\includegraphics[width=1.0\textwidth]{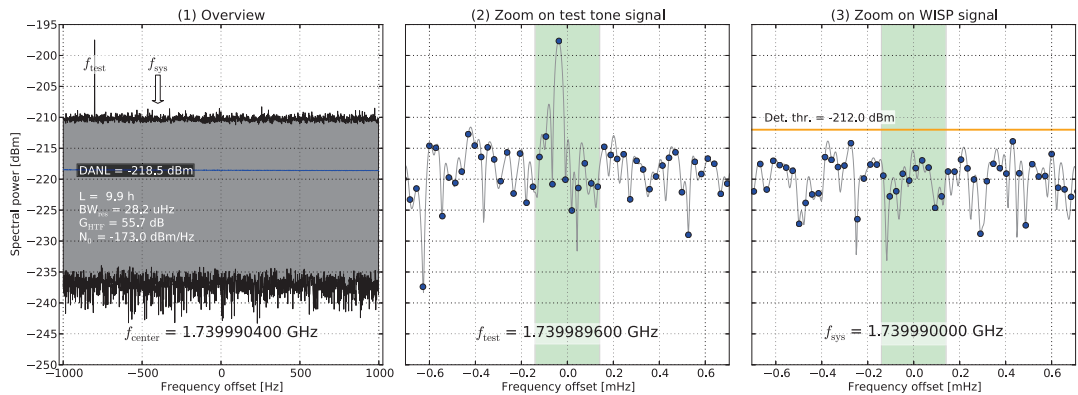}
\caption{Result of the ALP run in June 2013.(1) Spectrum showing the whole recorded frequency span. (2) Zoom on test tone signal. It is visible within the expected frequency range, marked in green. (3) Zoom on $f_{sys}$, no ALP signal is visible above the detection threshold.}
\end{center}
\end{figure}

\section{Conclusion and Outlook}
Profiting from the high mobility and ruggedness of a microwave based LSW experiment (compared to optical setups), we were able to successfully deploy and operate the apparatus within a 3~T MRI magnet. Over the course of one weekend sufficient experimental data was recorded to achieve comparable sensitivity to first generation optical LSW experiments (e.g., ALPS-1 at DESY).

Sensitivity could be further enhanced with a stronger magnet. There is a recent trend towards 7~T MRI systems in medical research, which might provide us an opportunity for a follow up ALPs measurement. Furthermore, sensitivity can be enhanced with larger and thus lower frequency cavities. One might think of a LSW setup consisting of 200~MHz SPS standing wave cavities \cite{src:SPS_SW_cav} within the 3~T M1 magnet \cite{src:M1_magnet} at CERN.  
\\
{\small
Special thanks to S.~W.~Rieger and the Brain \& Behaviour Laboratory of Geneva University, for making the MRI magnet accessible to us on weekends.
The authors would like to thank R.~Jones, E.~Jensen and the BE department management for encouragement and support.
Thanks to the organizers of the Patras Workshop for a very enjoyable and inspiring conference.
Supported by the Wolfgang-Gentner-Programme of the Bundesministerium f\"ur Bildung und Forschung
(BMBF).
}


\begin{footnotesize}

\end{footnotesize}


\end{document}